\documentclass[11pt,english,aps,manuscript,groupedaddress,superscriptaddress,txfonts,floatfix,showpacs,showkeys]{revtex4}
\usepackage[T1]{fontenc}
\usepackage[latin9]{inputenc}
\usepackage{array}
\usepackage{amsmath}
\usepackage{graphicx}
\usepackage{esint}

\makeatletter

\providecommand{\tabularnewline}{\\}

\@ifundefined{textcolor}{}
{%
 \definecolor{BLACK}{gray}{0}
 \definecolor{WHITE}{gray}{1}
 \definecolor{RED}{rgb}{1,0,0}
 \definecolor{GREEN}{rgb}{0,1,0}
 \definecolor{BLUE}{rgb}{0,0,1}
 \definecolor{CYAN}{cmyk}{1,0,0,0}
 \definecolor{MAGENTA}{cmyk}{0,1,0,0}
 \definecolor{YELLOW}{cmyk}{0,0,1,0}
}

\usepackage{dsfont}

\@ifundefined{showcaptionsetup}{}{%
 \PassOptionsToPackage{caption=false}{subfig}}
\usepackage{subfig}
\makeatother

\usepackage{babel}
\begin{document}

\title{Decays $B_{s}\rightarrow J/\psi+\eta$ and $B_{s}\rightarrow J/\psi+\eta'$
in the framework of covariant quark model}

\author{S. Dubni\v{c}ka}

\affiliation{Institute of Physics, Slovak Academy of Sciences, Bratislava, Slovak
Republic}

\author{A. Z. Dubni\v{c}kov\'a}

\affiliation{Dept. of Theoretical Physics, Comenius University, Bratislava, Slovak
Republic}

\author{M. A. Ivanov}

\affiliation{Bogoliubov Laboratory of Theoretical Physics, Joint Institute for
Nuclear Research, 141980 Dubna, Russia}

\author{A. Liptaj}

\affiliation{Institute of Physics, Slovak Academy of Sciences, Bratislava, Slovak
Republic}
\begin{abstract}
The covariant quark model represents an appropriate theoretical framework
to describe the recent results on $B_{s}\rightarrow J/\psi+\eta$
and $B_{s}\rightarrow J/\psi+\eta'$ decays from the Belle and LHCb
collaborations. In this article we present the main features of the
covariant quark model together with details on some of its aspects
and methods, which we consider to be important. Further we apply the
model specifically to the studied decay processes and give numerical
results on decay widths as they follow from the model. We conclude
that the model, with most of its parameters previously fixed from
different processes, is able to incorporate the new experimental measurements.
In particular, we found that the ratio of the branching fractions
of the decays $B_{s}$ into $J/\psi+\eta'$ and $J/\psi+\eta$ is
equal to $R\approx0.86$, in agreement with the data reported by Belle
and LHCb collaborations. 
\end{abstract}
\maketitle

\section{Introduction}

The low energy region of quantum chromodynamics (QCD) is an interesting
and challenging area of investigation nowadays. On one hand many accurate
data exists in this domain, especially results from hadron spectroscopy
and hadron decays, which might give a handle to understanding the
quark dynamics inside hadrons. Presently the precision and the amount
of these data continuously increases thanks to ongoing high-energy
experiments and heavy-quark factories. On the other hand theoretical
predictions encounter difficulties: since with decreasing energy the
coupling constant increases, the perturbative approach (pQCD) looses
its applicability and cannot provide valid results. Also lattice calculations
are generally seen as not yet enough developed and precise to be considered
as a well established low-energy solution of the QCD which would satisfactory
explain its numerous features. Undoubtedly however, some theoretical
description of hadron dynamics is needed at least for a correct description
of physic and background measured in particle detectors. One thus
has to rely on a model-dependent approach. 

In the wake of the recent measurement of the branching fractions of
$B_{s}$ meson decays into $J/\psi+\eta$ and $J/\psi+\eta'$ made
by Belle \cite{Belle:2012aa} and LHCb \cite{LHCb:2012cw} collaborations,
we perform the analysis of the above processes in the framework of
the covariant quark model.

Theoretical facets of these decays in the context of exploring CP
violation have been discussed in the papers Refs. \cite{Skands:2000ru,Colangelo:2010wg,DiDonato:2011kr,Fleischer:2011ib}.
Two points are important in analyzing: $SU_{f}(3)$ breaking and $\eta-\eta'$
mixing. There are attempts to include in addition the admixture of
glueball in the $\eta-\eta'$ mixing, see i.e. \cite{Liu:2012ib}.

The covariant quark model \cite{Branz:2009cd} is an ambitious model
with many appealing features. Its Lagrangian-based formulation leads
to a full Lorentz invariance and, in the limit of large number of
hadrons, it has only one free parameter per hadron. A distinctive
feature of this approach is that the multiquark states, such as baryons
(three quarks), tetraquarks (four quarks), etc., can be considered
and described as rigorously as the simplest quark-antiquark systems
(mesons). The model has wide application spectra and was already successfully
used to describe different types of heavy hadron decays and other
hadron-related observables. The last applications have been done for
studying the properties of the $B_{s}$ meson \cite{Ivanov:2011aa},
the light baryons \cite{Gutsche:2012ze}, the heavy $\Lambda_{b}$
baryon \cite{Gutsche:2013pp} and tetraquarks \cite{Dubnicka:2010kz,Dubnicka:2011mm}.
The results are reviewed in Ref. \cite{Ivanov:2013wy}.

In Sec. \ref{sec:Covariant-quark-model} we briefly discuss the basic
features of our approach. In Sec. \ref{sec:Model_features} we give
more details on the compositeness condition, the infrared confinement
and calculational techniques. The calculation of the matrix elements
of the decays $B_{s}\rightarrow J/\psi+\eta$ and $B_{s}\rightarrow J/\psi+\eta'$
is performed in Sec. \ref{sec:DecaysBs}. In Sec. \ref{sec:Results-and-discussion}
the calculated results are reported and comparison with available
experimental data is given. Finally, in Sec. \ref{sec:Summary} we
summarize our findings.

\section{Covariant quark model\label{sec:Covariant-quark-model}}

The interaction Lagrangian (density) of the model
\begin{equation}
\mathrm{\mathcal{L}_{int}=g_{H}H\left(x\right)\cdot J_{H}\left(x\right)+H.c.}\label{eq:Lagrangian}
\end{equation}
is constructed from the hadron field $\mathrm{H\left(x\right)}$ and
the quark current. The latter is in case of mesons written as
\begin{eqnarray}
\mathrm{J_{M}\left(x\right)} & \mathrm{=} & \mathrm{\int dx_{1}\int dx_{2}F_{M}\left(x;x_{1},x_{2}\right)\times\bar{q}_{2}^{a}\left(x_{2}\right)\Gamma_{M}q_{1}^{a}\left(x_{1}\right)}
\end{eqnarray}
and makes appear that, within the model, the interaction is mediated
only by the quarks and the gluons are absent. The form of the vertex
function $\mathrm{F_{M}}$ is chosen such as to reflect the intuitive
expectations about relative quark-hadron positions
\begin{eqnarray}
\mathrm{F_{M}\left(x;x_{1},x_{2}\right)} & \mathrm{=} & \mathrm{\delta^{\left(4\right)}\left(x-w_{1}x_{1}-w_{2}x_{2}\right)\times\Phi_{M}\left[\left(x_{1}-x_{2}\right)^{2}\right],}
\end{eqnarray}
where we require $\mathrm{w_{1}+w_{2}=1}$. We actually adopt the
most natural choice
\begin{equation}
\mathrm{w_{i}=\frac{m_{i}}{m_{1}+m_{2}},\: i=1,2}\label{eq:weights}
\end{equation}
where the barycenter of the hadron is identified with the barycenter
of the quark system. The interaction strength $\mathrm{\Phi_{M}\left[\left(x_{1}-x_{2}\right)^{2}\right]}$
is assumed to have a Gaussian form which is in the momentum representation
written as
\begin{equation}
\mathrm{\widetilde{\Phi}_{M}\left(-p^{2}\right)=\exp\left(\frac{p^{2}}{\Lambda_{M}^{2}}\right)}.
\end{equation}
Here $\mathrm{\Lambda_{M}}$ is a hadron-related size parameter which
is regarded as a free parameter of the model. Additional free parameters
are the quark masses
\begin{eqnarray}
\mathrm{m_{u,d}=0.235\: GeV,} &  & \mathrm{m_{s}=0.424\: GeV,}\nonumber \\
\mathrm{m_{c}=2.16\: GeV,} &  & \mathrm{m_{b}=5.09\: GeV}
\end{eqnarray}
and a universal cut-off parameter $\mathrm{\lambda_{cut-off}}$, which
is commented on in more details later in this text. The numerical
values of these parameters as well as of size parameters for several
hadrons were fixed by fitting the model to measurement data of simple
quantities, such as leptonic decay constants and electromagnetic decay
widths \cite{Ivanov:2011aa}. The value of the size parameter for
$B_{s}$ meson is $\mathrm{\Lambda_{B_{s}}=1.95\: GeV}$. The model
thus has in total $\mathrm{N_{H}+5}$ parameters: for each of $\mathrm{N_{H}}$
hadrons one $\Lambda$ parameter, four quark masses and one universal
cut-off. The coupling constants $\mathrm{g_{M}}$ can be related to
these parameters using the so-called \textit{compositeness condition},
which is discussed in the following section.

\section{Compositeness condition, calculation methods and infrared confinement\label{sec:Model_features}}

In this section we give more details about some issues related to
the model. As announced previously, one of them is the compositeness
condition.

The quark fields as well as the hadron field enter the interaction
Lagrangian of the model (\ref{eq:Lagrangian}) as elementary although,
in nature, the hadrons are compound of quarks. An effort was made
in theoretical physics to find an appropriate description of composite
particles. The authors of the references \cite{Salam:1962ap,Weinberg:1962hj}
argue, that the hadron fields renormalization constant $\mathrm{Z_{H}^{\frac{1}{2}}}$
can be interpreted as the matrix element between the physical state
and the corresponding bare state. The case $\mathrm{Z_{H}=0}$ thus
corresponds to a state not containing the bare state and can be therefore
properly interpreted as a bound state. This idea was used and introduced
into the covariant quark model \cite{Efimov:1993zg,Efimov:1988yd}.
The renormalization constant is expressed through the derivative of
the meson mass operator and takes the form
\begin{equation}
\mathrm{Z_{M}=1-\frac{3g_{M}^{2}}{4\pi^{2}}\tilde{\Pi}_{M}^{'}\left(m_{M}^{2}\right)=0}.
\end{equation}
The coupling constants are in this way eliminated as free parameters.
This not only gives the model more predictive power but also helps
to stabilize model predictions over wide spectra of hadron data. The
first step in calculation of physical observables is thus the determination
of couplings $\mathrm{g_{M}}$ of participating hadrons.

In the case of the pseudoscalar and vector mesons the derivative of
the meson mass operator can be calculated in the following way
\begin{eqnarray*}
\mathrm{\widetilde{\Pi}'_{P}(p^{2})} & \mathrm{=} & \mathrm{\frac{1}{2p^{2}}\, p^{\alpha}\frac{d}{dp^{\alpha}}\,\int\!\!\frac{d^{4}k}{4\pi^{2}i}\,\widetilde{\Phi}_{P}^{2}(-k^{2})\times{\rm tr}\biggl[\gamma^{5}S_{1}(k+w_{1}p)\gamma^{5}S_{2}(k-w_{2}p)\biggr]}\mathrm{}\\
 & \mathrm{=} & \mathrm{\frac{1}{2p^{2}}\,\int\!\!\frac{d^{4}k}{4\pi^{2}i}\,\widetilde{\Phi}_{P}^{2}(-k^{2})\times\Big\{ w_{1}\,{\rm tr}\biggl[\gamma^{5}S_{1}(k+w_{1}p)\!\not\! pS_{1}(k+w_{1}p)\gamma^{5}S_{2}(k-w_{2}p)\biggr]}\\
 &  & \mathrm{-w_{2}\,{\rm tr}\biggl[\gamma^{5}S_{1}(k+w_{1}p)\gamma^{5}S_{2}(k-w_{2}p)\!\not\! p\, S_{2}(k-w_{2}p)\biggr]\Big\},}
\end{eqnarray*}
\begin{eqnarray}
\mathrm{\widetilde{\Pi}'_{V}(p^{2})} & \mathrm{=} & \mathrm{\frac{1}{3}\left(g_{\mu\nu}-\frac{p_{\mu}p_{\nu}}{p^{2}}\right)\times\frac{1}{2p^{2}}\, p^{\alpha}\frac{d}{dp^{\alpha}}\,\int\!\!\frac{d^{4}k}{4\pi^{2}i}\,\widetilde{\Phi}_{V}^{2}(-k^{2})\times{\rm tr}\biggl[\gamma^{\mu}S_{1}(k+w_{1}p)\gamma^{\nu}S_{2}(k-w_{2}p)\biggr]}\nonumber \\
 & \mathrm{=} & \mathrm{\frac{1}{3}\left(g_{\mu\nu}-\frac{p_{\mu}p_{\nu}}{p^{2}}\right)\frac{1}{2p^{2}}\,\int\!\!\frac{d^{4}k}{4\pi^{2}i}\times\widetilde{\Phi}_{V}^{2}(-k^{2})}\nonumber \\
 &  & \mathrm{\times\Big\{ w_{1}\,{\rm tr}\biggl[\gamma^{\mu}S_{1}(k+w_{1}p)\not\! p\: S_{1}(k+w_{1}p)\gamma^{\nu}S_{2}(k-w_{2}p)\biggr]}\nonumber \\
 &  & \mathrm{-w_{2}\,{\rm tr}\biggl[\gamma^{\mu}S_{1}(k+w_{1}p)\gamma^{\nu}S_{2}(k-w_{2}p)\!\not\! p\, S_{2}(k-w_{2}p)\biggr]\Big\},}
\end{eqnarray}
where $\mathrm{\widetilde{\Phi}_{M}\left(-k^{2}\right)}$ is the Fourier
transform of the vertex function $\mathrm{\Phi_{M}\left[\left(x_{1}-x_{2}\right)^{2}\right]}$
and $\mathrm{S_{i}(k)}$ is the quark propagator. We have used free
fermion propagators for the quarks given by 
\begin{equation}
\mathrm{S_{q}(k)=\frac{1}{m_{q}-\not\! k}}
\end{equation}
 with an effective constituent quark mass $\mathrm{m_{q}}$. 

In computation of Feynman diagrams we use, in the momentum space,
the Schwinger representation of the quark propagator
\begin{eqnarray}
\mathrm{S_{q}(k)} & \mathrm{=} & \mathrm{\frac{m_{q}+\not\! k}{m_{q}^{2}-k^{2}}}\nonumber \\
 & \mathrm{=} & \mathrm{\left(m_{q}+\not\! k\right)\intop_{0}^{\infty}d\alpha\: e^{-\alpha\left(m^{2}-k^{2}\right)}}
\end{eqnarray}
The general form of a resulting Feynman diagrams is
\begin{equation}
\mathrm{\Pi\left(p_{1},\ldots,p_{m}\right)=\intop_{0}^{\infty}d^{n}\alpha\int\left[d^{4}k\right]^{\ell}\Phi\times\exp\left\{ -\sum_{i=1}^{n}\alpha_{i}\left[m_{i}^{2}-\left(K_{i}+P_{i}\right)^{2}\right]\right\} ,}
\end{equation}
where $\mathrm{K_{i}}$ represents a linear combination of loop momenta,
$\mathrm{P_{i}}$ stands for a linear combination of external momenta
and $\mathrm{\Phi}$ refers to the numerator product of propagators
and vertex functions. The evaluation of these expressions can be much
simplified if done in a smart way. One can use two operator identities,
of which the first one
\begin{align}
\mathrm{\int d^{4}k\: P\left(k\right)e^{2kr}} & \mathrm{=}\mathrm{\int d^{4}k\: P\left(\frac{1}{2}\frac{\partial}{\partial r}\right)e^{2kr}}\nonumber \\
 & \mathrm{=}\mathrm{P\left(\frac{1}{2}\frac{\partial}{\partial r}\right)\int d^{4}k\: e^{2kr}}
\end{align}
is suited for en elegant loop momenta integration. The second one
\begin{equation}
\mathrm{\intop_{0}^{\infty}d^{n}\alpha\: P\left(\frac{1}{2}\frac{\partial}{\partial r}\right)e^{-\frac{r^{2}}{a}}=\intop_{0}^{\infty}d^{n}\alpha\: e^{-\frac{r^{2}}{a}}P\left(\frac{1}{2}\frac{\partial}{\partial r}-\frac{r}{a}\right)\mathds{1},}
\end{equation}
where $\mathrm{r=r\left(\alpha_{i}\right)}$ and $\mathrm{a=a\left(\Lambda_{M},\alpha_{i}\right)}$,
simplifies the computation following the trace evaluation: the polynomial
in the derivative operator which results from the trace can be applied
to an identity, instead being applied to a more complicated exponential
function.

The last point which remains to be discussed is the cut-off we apply
in the integration over the Schwinger parameters. This integration
is multidimensional with the limits from $0$ to $+\infty$. In order
to arrive to a single cut-off parameter we firstly transform the integral
over an infinite space into an integral over a simplex convoluted
with only one-dimensional improper integral. For that purpose we use
the $\delta$-function form of the identity
\begin{equation}
\mathrm{\mathds{1}=\intop_{0}^{\infty}dt\:\delta\left(t-\sum_{i=1}^{n}\alpha_{i}\right)}
\end{equation}
from which follows
\begin{equation}
\Pi=\intop_{0}^{\infty}dt\: t^{n-1}\intop_{0}^{1}d^{n}\alpha\:\delta\left(t-\sum_{i=1}^{n}\alpha_{i}\right)\times W\left(t\alpha_{1},\ldots,t\alpha_{1}\right),
\end{equation}
where $W$ represents the integrand of Schwinger parameters. The cut-off
$\lambda$ is then introduced in a natural way
\begin{equation}
\intop_{0}^{\infty}dt\: t^{n-1}\ldots\rightarrow\intop_{0}^{1/\lambda^{2}}dt\: t^{n-1}\ldots.
\end{equation}
Such a cut-off makes the integral to be an analytic function without
any singularities. In this way all potential thresholds in the quark
loop diagrams are removed together with corresponding branch points
\cite{Branz:2009cd}. Within covariant quark model the cut-off parameter
is universal for all processes and its value, as obtained from a fit
to data, is
\begin{equation}
\mathrm{\lambda_{cut-off}=0.181\: GeV.}
\end{equation}
The integrals are computed numerically.

\section{Decays $B_{s}\rightarrow J/\psi+\eta$ and $B_{s}\rightarrow J/\psi+\eta'$
\label{sec:DecaysBs}}

\begin{figure*}
\subfloat[]{\begin{centering}
\includegraphics[width=0.3\linewidth]{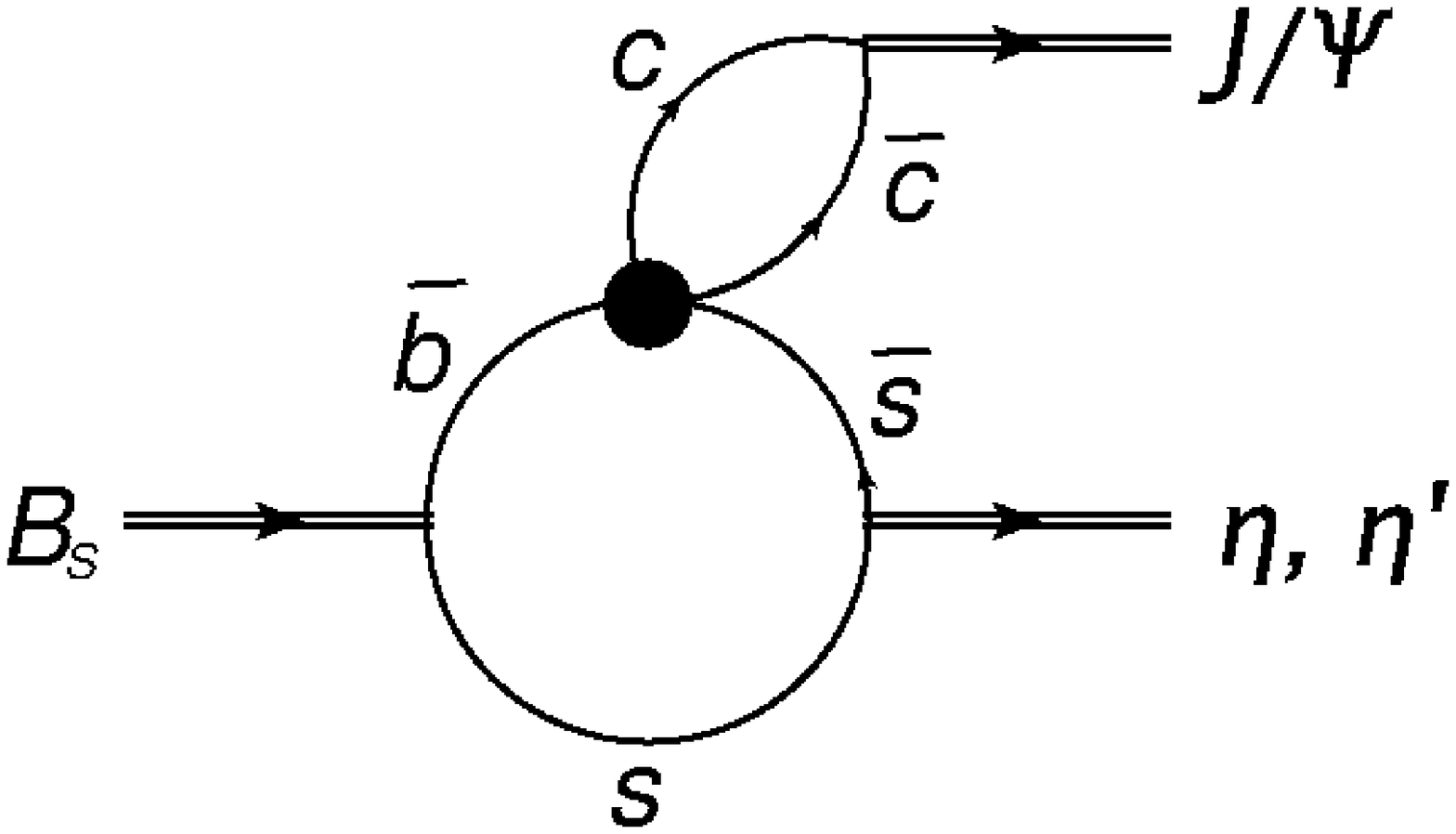}\label{SubFigA}
\par\end{centering}

}\subfloat[]{\centering{}\includegraphics[width=0.3\linewidth]{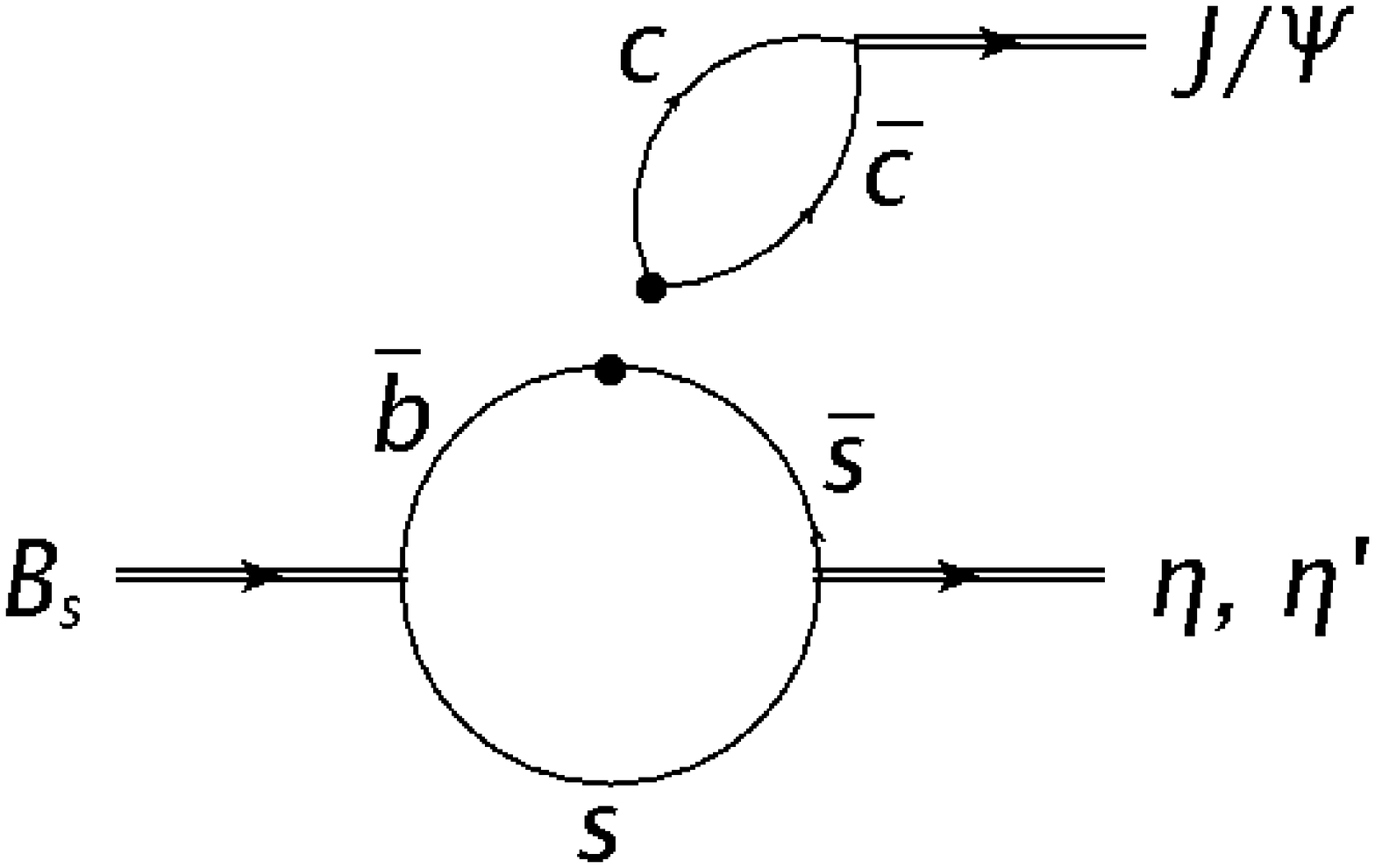}\label{SubFigB}}\caption{a) Diagram of $B_{s}\rightarrow J/\psi+\eta^{\left(\prime\right)}$
decay. b) Factorization of the diagram.}

\end{figure*}
The diagram of the $B_{s}\rightarrow J/\psi+\eta^{\left(\prime\right)}$
decay is shown in Fig. \ref{SubFigA}. The $\eta$ and $\eta'$ particles
are mixture of the light and the $s$-quark component. In the approximation
$m_{u}=m_{d}\equiv m_{q}$ one can write their quark content as
\begin{eqnarray}
\mathrm{\mathrm{\eta}} & \mathrm{\mathrm{:}} & \mathrm{\mathrm{-\frac{1}{\sqrt{2}}\sin\delta\left(u\bar{u}+d\bar{d}\right)-\cos\delta\left(s\bar{s}\right)}}\nonumber \\
 &  & \mathrm{\mathrm{=-\sin\delta\left(q\bar{q}\right)-\cos\delta\left(s\bar{s}\right),}}
\end{eqnarray}
\begin{eqnarray}
\mathrm{\mathrm{\eta'}} & \mathrm{:} & \frac{1}{\sqrt{2}}\mathrm{\cos\delta\left(u\bar{u}+d\bar{d}\right)-\sin\delta\left(s\bar{s}\right)}\nonumber \\
 &  & \mathrm{=\cos\delta\left(q\bar{q}\right)-\sin\delta\left(s\bar{s}\right),}
\end{eqnarray}
with 
\begin{equation}
\mathrm{q\bar{q}=\frac{1}{\sqrt{2}}\left(u\bar{u}+d\bar{d}\right)}
\end{equation}
and 
\begin{eqnarray}
\mathrm{\delta} & \mathrm{=} & \mathrm{\theta_{P}-\theta_{I},}\nonumber \\
 &  & \mathrm{\theta_{I}=\arctan\frac{1}{\sqrt{2}}\approx35.26^{\circ},}\nonumber \\
 &  & \theta_{P}\approx-13.34^{\circ}.
\end{eqnarray}
The value of the pseudoscalar angle $\theta_{P}$ is deduced from
the Ref. \cite{Ambrosino:2006gk}, where however a different convention
is used for the mixing angle $\mathrm{\varphi_{P}}$. It can be related
to ours by $\mathrm{\delta=\varphi_{P}-\pi/2}$.

We describe the decay only through the dominant $s$-quark contribution
to the $\eta^{\left(\prime\right)}$ meson since the light quark one
results from higher-order diagrams. The quark-hadron interaction is
described by the Lagrangians from the Section \ref{sec:Covariant-quark-model}
(see Appendix \ref{sub:Interaction-lagrangians-in}) and we use an
effective theory with four-quark interaction to describe the production
of the $J/\psi$ particle. The Lagrangian of this interaction is written
as
\begin{equation}
\mathrm{\mathcal{L}_{eff}=\frac{G_{F}}{\sqrt{2}}V_{cb}V_{cs}^{*}\sum_{i}C_{i}Q_{i},}
\end{equation}
where $\mathrm{G_{F}}$ is the Fermi Coupling Constant, $\mathrm{V_{xy}}$
refers to the elements of the CKM matrix, $\mathrm{C_{i}}$ are the
Wilson coefficients (Ref. \cite{Altmannshofer:2008dz}) and $Q_{i}$
are operators listed in Appendix \ref{sub:Four-quark-vertex-operators}.
To each component of both $\eta$ and $\eta'$ particles we associate
a size parameter. So, in general, the description requires four free
parameters $\mathrm{\Lambda_{\eta}^{q\bar{q}}}$, $\mathrm{\Lambda_{\eta}^{s\bar{s}}}$
, $\mathrm{\Lambda_{\eta'}^{q\bar{q}}}$ and $\mathrm{\Lambda_{\eta'}^{s\bar{s}}}$.
We consider the mixing angle $\mathrm{\delta}$ as fixed although,
in principle, it also might be varied to determine its most suited
value within the covariant quark model. As a model-independent parameter
it can be compared to other models and to data.

An important simplification in calculations comes from the evaluation
of the $\mathrm{Q_{i}}$ operator matrix elements. In can be shown
that the expression factorizes into a part that corresponds to the
$\mathrm{B_{s}\rightarrow\eta^{\left(\prime\right)}}$ transition
form factor and a part that is proportional to the decay constant
of the $\mathrm{J/\psi}$ particle (Fig \ref{SubFigB} and Appendix
\ref{sub:Some-elements-of}). 

It is readily seen from the Fig. \ref{SubFigB} that only the strange
quark component gives the contribution to the color-suppressed $B_{s}\rightarrow\eta^{\left(\prime\right)}$
decays. The invariant matrix element describing this decay is written
as
\begin{equation}
\mathrm{M\left[B_{s}(p_{1})\to P_{s\bar{s}}(p_{2})+J/\psi(q)\right]=\frac{G_{F}}{\sqrt{2}}V_{cb}V_{cs}^{\dagger}C_{W}F_{+}^{B_{s}P_{s\bar{s}}}(q^{2})\, m_{J/\psi}f_{J/\psi}\,(p_{1}+p_{2})\cdot\epsilon_{J/\psi},}
\end{equation}
where $\mathrm{q=p_{1}-p_{2}}$, $\mathrm{q\cdot\epsilon_{J/\psi}=0}$,
$\mathrm{q^{2}=m_{J/\psi}^{2}}$ and the color factor $\mathrm{C_{W}}=\mathrm{C_{1}}+\xi\mathrm{C_{2}}+\mathrm{C_{3}}+\xi\mathrm{C_{4}}+\mathrm{C_{5}}+\xi\mathrm{C_{6}}$
with $\mathrm{\xi=1/N_{c}}$. The terms multiplied by the color factor
$\mathrm{\xi}$ will be dropped in the numerical calculations according
to the $\mathrm{1/N_{c}}$ expansion.

The leptonic decay constants of the pseudoscalar and vector mesons
are defined by
\begin{eqnarray}
\mathrm{N_{c}\, g_{P}\!\int\!\!\frac{d^{4}k}{(2\pi)^{4}i}\,\widetilde{\Phi}_{P}(-k^{2})\,{\rm tr}\biggl[O^{\,\mu}S_{1}(k+w_{1}p)\gamma^{5}S_{2}(k-w_{2}p)\biggr]} & \mathrm{=} & \mathrm{f_{P}p^{\mu},\quad p^{2}=m_{P}^{2},}\nonumber \\
\mathrm{N_{c}\, g_{V}\!\int\!\!\frac{d^{4}k}{(2\pi)^{4}i}\,\widetilde{\Phi}_{V}(-k^{2})\,{\rm tr}\biggl[O^{\,\mu}S_{1}(k+w_{1}p)\not\!\epsilon_{V}S_{2}(k-w_{2}p)\biggr]} & \mathrm{=} & \mathrm{m_{V}f_{V}\epsilon_{V}^{\mu},\quad p^{2}=m_{V}^{2}.}
\end{eqnarray}

The form factors of the $\mathrm{P_{[\bar{q}_{3}q_{2}]}-P_{[\bar{q}_{1}q_{3}]}^{\,\prime}}$
transition are given by
\begin{eqnarray}
\mathrm{\langle P_{[\bar{q}_{1}q_{3}]}^{\,\prime}(p_{2})\,|\,\bar{q}_{2}\, O^{\,\mu}\, q_{1}\,|P_{[\bar{q}_{3}q_{2}]}(p_{1})\rangle} & \mathrm{=} & \mathrm{N_{c}\, g_{P}\, g_{P^{\,'}}\!\int\!\frac{d^{4}k}{(2\pi)^{4}i}\:\widetilde{\Phi}_{P}\Big(-(k+w_{13}p_{1})^{2}\Big)\,\widetilde{\Phi}_{P^{\,'}}\Big(-(k+w_{23}p_{2})^{2}\Big)}\nonumber \\
 &  & \qquad\mathrm{\times{\rm tr}\biggl[O^{\,\mu}\, S_{1}(k+p_{1})\,\gamma^{5}\, S_{3}(k)\,\gamma^{5}\, S_{2}(k+p_{2})\biggr]}\nonumber \\
 & \mathrm{=} & \mathrm{F_{+}(q^{2})\, P^{\,\mu}+F_{-}(q^{2})\, q^{\,\mu}}
\end{eqnarray}
where in addition to Eq. (\ref{eq:weights}) we have introduced a
two-subscript notation $\mathrm{w_{ij}=m_{j}/(m_{i}+m_{j})}$ such
that $\mathrm{w_{ij}+w_{ji}=1}$. The calculation of the widths is
straightforward. One has
\[
\mathrm{\Gamma(B_{s}\to J/\psi+\eta)=\frac{G_{F}^{2}}{4\pi}|V_{cb}V_{cs}^{\dagger}|^{2}C_{W}^{2}\, f_{J/\psi}^{2}\,|\mathbf{q}_{\eta}|^{3}\cos^{2}\delta\left(F_{+}^{B_{s}\eta}(m_{J/\psi}^{2})\right)^{2},}
\]
\begin{equation}
\mathrm{\Gamma(B_{s}\to J/\psi+\eta')=\frac{G_{F}^{2}}{4\pi}|V_{cb}V_{cs}^{\dagger}|^{2}C_{W}^{2}\, f_{J/\psi}^{2}\,|\mathbf{q}_{\eta'}|^{3}\sin^{2}\delta\left(F_{+}^{B_{s}\eta'}(m_{J/\psi}^{2})\right)^{2},}\label{eq:Gamma_Bs}
\end{equation}
where $\mathrm{|{\mathbf{q}_{P}}|=\lambda^{1/2}(m_{B_{s}}^{2},m_{P}^{2},m_{J/\psi}^{2})/(2m_{B_{s}})}$
is the momentum of the outgoing particles in the rest frame of the
decaying particle.

In addition to the Belle results, we need further data in order to
over-constrain the model. We have chosen the following decays
\begin{eqnarray}
\eta\rightarrow\gamma\gamma & \qquad & \varphi\rightarrow\eta\gamma\nonumber \\
\eta'\rightarrow\gamma\gamma & \qquad & \varphi\rightarrow\eta'\gamma\nonumber \\
\rho^{0}\rightarrow\eta\gamma & \qquad & B_{d}\rightarrow J/\psi+\eta\nonumber \\
\omega\rightarrow\eta\gamma & \qquad & B_{d}\rightarrow J/\psi+\eta'\nonumber \\
\eta'\rightarrow\omega\gamma & \qquad
\end{eqnarray}
These processes have already been previously described by the covariant
quark model \cite{Branz:2009cd} and can be straightforwardly included
into a global fit.

\section{Results and discussion\label{sec:Results-and-discussion}}

\begin{table*}
\begin{centering}
\begin{tabular}{>{\centering}p{2cm}>{\centering}p{2.5cm}>{\centering}p{3.5cm}>{\centering}p{0.5cm}>{\centering}p{2cm}>{\centering}p{2cm}>{\centering}p{2cm}}
\hline 
Observable & Covariant  & \multicolumn{2}{>{\centering}b{3.5cm}}{Experiment } & \multicolumn{3}{c}{Other models}\tabularnewline
 & quark model & \multicolumn{2}{c}{} & M\"unz \cite{Munz:1993si} & Becchi \cite{Becchi:1965zza} & Jaus \cite{Jaus:1991cy}\tabularnewline
\hline 
$\Gamma_{\eta\rightarrow\gamma\gamma}$ & $\mathrm{0.380\: keV}$ & $\mathrm{0.511\pm0.028\: keV}$ & \cite{Babusci:2012ik} & $\mathrm{0.440\: keV}$ & - & $\mathrm{0.485\: keV}$\tabularnewline
$\mathrm{\Gamma_{\eta'\rightarrow\gamma\gamma}}$ & $\mathrm{3.74\: keV}$ & $\mathrm{4.34\pm0.14\: keV}$ & \cite{Beringer:1900zz} & $\mathrm{2.90\: keV}$ & - & $\mathrm{2.90\: keV}$\tabularnewline
$\mathrm{\Gamma_{\rho^{0}\rightarrow\eta\gamma}}$ & $\mathrm{53.07\: keV}$ & $\mathrm{44.73\pm2.99\: keV}$ & \cite{Beringer:1900zz} & - & $\mathrm{44\: keV}$ & $\mathrm{59\: keV}$\tabularnewline
$\mathrm{\Gamma_{\omega\rightarrow\eta\gamma}}$ & $\mathrm{6.21\: keV}$ & $\mathrm{3.91\pm0.34\: keV}$ & \cite{Beringer:1900zz} & - & $\mathrm{6.4\: keV}$ & $\mathrm{8.7\: keV}$\tabularnewline
$\mathrm{\Gamma_{\eta'\rightarrow\omega\gamma}}$ & $\mathrm{9.49\: keV}$ & $\mathrm{5.47\pm0.63\: keV}$ & \cite{Beringer:1900zz} & - & - & $\mathrm{4.8\: keV}$\tabularnewline
$\mathrm{\Gamma_{\varphi\rightarrow\eta\gamma}}$ & $\mathrm{42.59\: keV}$ & $\mathrm{58.90\pm2.45\: keV}$ & \cite{Beringer:1900zz} & - & $\mathrm{304\: keV}$ & $\mathrm{55.3\: keV}$\tabularnewline
$\mathrm{\Gamma_{\varphi\rightarrow\eta'\gamma}}$ & $\mathrm{0.276\: keV}$ & $\mathrm{0.281\pm0.015\: keV}$ & \cite{Beringer:1900zz} & - & - & $\mathrm{0.57\: keV}$\tabularnewline
$\mathrm{{\cal B}_{B_{d}\rightarrow J/\psi+\eta}}$ & $\mathrm{16.5\times10^{-6}}$ & $\mathrm{\left(12.3\pm1.9\right)\times10^{-6}}$ & \cite{Chang:2012gnb} & - & - & -\tabularnewline
$\mathrm{{\cal B}_{B_{d}\rightarrow J/\psi+\eta'}}$ & $\mathrm{12.2\times10^{-6}}$ & $\mathrm{<7.4\times10^{-6}}$ & \cite{Chang:2012gnb} & - & - & -\tabularnewline
$\mathrm{{\cal B}_{B_{s}\rightarrow J/\psi+\eta}}$ & $\mathrm{4.67\times10^{-4}}$ & $\mathrm{\left(5.10\pm1.12\right)\times10^{-4}}$ & \cite{Belle:2012aa} & - & - & -\tabularnewline
$\mathrm{{\cal B}_{B_{s}\rightarrow J/\psi+\eta'}}$ & $\mathrm{4.04\times10^{-4}}$ & $\mathrm{\left(3.71\pm0.95\right)\times10^{-4}}$ & \cite{Belle:2012aa} & - & - & - \tabularnewline
\hline 
\end{tabular}
\par\end{centering}

\caption{Decay widths and branching fractions for selected processes: model
predictions and data.}
\label{TAB:results}
\end{table*}
The optimal model parameters (in $\mathrm{GeV}$)
\begin{eqnarray}
 & \mathrm{\Lambda_{\eta}^{q\bar{q}}=0.881,\quad\Lambda_{\eta}^{s\bar{s}}=1.973}\nonumber \\
 & \mathrm{\Lambda_{\eta'}^{q\bar{q}}=0.257,\quad\Lambda_{\eta'}^{s\bar{s}}=2.797}
\end{eqnarray}
were obtained from a $\mathrm{\chi^{2}}$ fit to the data. The corresponding
description of the data by the covariant quark model is shown in Table
\ref{TAB:results}.

First, we would like to discuss the ratio of the branching fractions
of $B_{s}\to J/\psi+\eta'$ and $B_{s}\to J/\psi+\eta$ decays which
has been measured by both, Belle \cite{Belle:2012aa} and LHCb \cite{LHCb:2012cw}
collaborations. They reported the following values for this ratio
\begin{eqnarray}
R & \equiv & \frac{\Gamma(B_{s}\to J/\psi+\eta')}{\Gamma(B_{s}\to J/\psi+\eta)}=\begin{cases}
0.73\pm0.14\pm0.02 & \mathrm{Belle}\;[1]\\
0.90\pm0.09_{-0.02}^{+0.06} & \mathrm{LHCb}\;[2]
\end{cases}
\end{eqnarray}
As follows from the Eq. \ref{eq:Gamma_Bs} the ratio $R$ is defined
by
\begin{equation}
R^{\,{\rm theor}}=\underbrace{\frac{|\mathbf{q}_{\eta'}|^{3}}{|\mathbf{q}_{\eta}|^{3}}\tan^{2}\delta}_{\approx1.04}\times\underbrace{\left(\frac{F_{+}^{B_{s}\eta'}}{F_{+}^{B_{s}\eta}}\right)^{2}}_{\approx0.83}\approx0.86.
\end{equation}
One can see that the nontrivial dependence of the form factor $\mathrm{F_{+}}$
on the $\eta$ and $\eta'$ masses and size parameters provides a
reduction of the model independent result $1.04$ to $0.86$.

The interpretation of the results might by done in several steps.
Firstly, one observes that all model predictions are of the order
of the experimental numbers. Majority of them have the relative error
smaller then 30\%, when compared to the data. The most important relative
error is 73\% in case of $\mathrm{\Gamma_{\eta'\rightarrow\omega\gamma}}$,
still significantly smaller then factor two. The latter case actually
suggests one might want to consider the possibility of a gluonium
content of the $\eta^{\left(\prime\right)}$ meson, as discussed in
the Ref. \cite{Ambrosino:2006gk} (and other references therein).
On the other hand one must admit that, when the experimental errors
are taken into account, the model prediction are usually quite outside
the error intervals. Here one can argue in two ways. Firstly, the
model is only an approximation to the first-principle theory. One
thus, from beginning, does not expect the model to be fully accurate.
Consequently, the goodness of the model when interpreted through the
data error intervals depends on the data precision measurement. Every
approximate model becomes ``very bad'' when the data becomes very
precise. So the point of view based on the error intervals might not
be the most suited one.

An optional criterion might be a comparison to the experimental needs.
Experiments usually need a model that allows for an appropriate correction
of detector effects. In fact they usually need more models so as to
be able to establish a systematic error related to the data correction.
We think that within this logic, the results of the covariant quark
model make the model fully acceptable and legitimate is development
and eventual use.

Yet, an additional approach to rate the model is to compare it to
other existing models \cite{Jaus:1991cy,Munz:1993si,Becchi:1965zza,Dorokhov:2011zf}.
We have chosen some of not very numerous works, that give explicit
numbers for a set of observables overlapping with those chosen by
us. When comparing processes in common (Table \ref{TAB:results}),
none of this models describes the data better than ours (if total
the $\mathrm{\chi^{2}}$ is calculated). This fact confirms, that
the covariant quark model is a very competitive one among the available
models.

Finally, one can still await further confirmation and more precision
in the experimental data, especially in the case of recent results
obtained by a single collaboration which have not yet been independently
cross-checked. The final picture concerning the data description by
the model might still change.

\section{Summary\label{sec:Summary}}

We have calculated the matrix elements and branching fractions of
the decays of $B_{s}$ into $J/\psi+\eta'$ and $J/\psi+\eta$ in
the framework of the covariant quark model. We have used the model
parameters which have been fixed in our previous papers except the
size parameters characterizing the distributions of non-strange and
strange quarks within the $\eta$ and $\eta'$. We fix them by fitting
the available experimental data on two-body electromagnetic decays
involving the $\eta$ and $\eta'$ and the above $B_{s}$ decay. In
particular, we have found that the ratio of the branching fractions
of the decays $B_{s}$ into $J/\psi+\eta'$ and $J/\psi+\eta$ is
equal to $R\approx0.86$ in agreement with the data reported by Belle
and LHCb collaborations. 
\begin{acknowledgments}
The work was partly supported by Slovak Grant Agency for Sciences
VEGA, grant No. 2/0009/10 (S. Dubni\v{c}ka, A. Z. Dubni\v{c}kov\'a,
A. Liptaj) and Joint research project of Institute of Physics, SAS
and Bogoliubov Laboratory of Theoretical Physics, JINR, No. 01-3-1070
(S. Dubni\v{c}ka, A. Z. Dubni\v{c}kov\'a, M. A. Ivanov and A. Liptaj).
\end{acknowledgments}
\appendix

\section{Expressions and Formulas}

We make use of this Appendix to display longer formulas referred in
the text.

\subsection{Lagrangian of the model\label{sub:Interaction-lagrangians-in}}

The Lagrangian (density) is written as
\[
\mathrm{\mathcal{L}=\mathcal{L}_{B_{S}}+\mathcal{L}_{\eta}+\mathcal{L}_{J/\psi}+\mathcal{L}_{eff}}
\]
with $\mathrm{\mathcal{L}_{eff}}$ being given in the text.
\[
\mathrm{\mathcal{L}_{B_{S}}\left(x\right)=g_{B_{S}}\overline{B}_{S}^{0}\left(x\right)\iint dx_{1}dx_{2}\delta\left(x-w_{b}x_{1}-w_{s}x_{2}\right)\phi_{B_{S}}\left[\left(x_{1}-x_{2}\right)^{2}\right]\bar{b}\left(x_{1}\right)i\gamma^{5}s\left(x_{2}\right)\:+\: h.\: c.}
\]
\begin{multline*}
\mathrm{\mathcal{L}_{\eta}\left(x\right)=g_{\eta}\eta\left(x\right)\iint dx_{1}dx_{2}\delta\left(x-\frac{1}{2}x_{1}-\frac{1}{2}x_{2}\right)\phi_{\eta}\left[\left(x_{1}-x_{2}\right)^{2}\right]}\\
\mathrm{\times\left\{ -\frac{1}{\sqrt{2}}\sin\left(\delta\right)\left[\bar{u}\left(x_{1}\right)i\gamma^{5}u\left(x_{2}\right)+\bar{d}\left(x_{1}\right)i\gamma^{5}d\left(x_{2}\right)\right]-\cos\left(\delta\right)\left[\bar{s}\left(x_{1}\right)i\gamma^{5}s\left(x_{2}\right)\right]\right\} }
\end{multline*}
\begin{multline*}
\mathrm{\mathcal{L}_{\eta'}\left(x\right)=g_{\eta'}\eta'\left(x\right)\iint dx_{1}dx_{2}\delta\left(x-\frac{1}{2}x_{1}-\frac{1}{2}x_{2}\right)\phi_{\eta'}\left[\left(x_{1}-x_{2}\right)^{2}\right]}\\
\mathrm{\times\left\{ \frac{1}{\sqrt{2}}\cos\left(\delta\right)\left[\bar{u}\left(x_{1}\right)i\gamma^{5}u\left(x_{2}\right)+\bar{d}\left(x_{1}\right)i\gamma^{5}d\left(x_{2}\right)\right]-\sin\left(\delta\right)\left[\bar{s}\left(x_{1}\right)i\gamma^{5}s\left(x_{2}\right)\right]\right\} }
\end{multline*}
\[
\mathrm{\mathcal{L}_{J/\psi}=g_{\psi}\psi_{\mu}\left(x\right)\iint dx_{1}dx_{2}\delta\left(x-\frac{1}{2}x_{1}-\frac{1}{2}x_{2}\right)\phi_{\psi}\left[\left(x_{1}-x_{2}\right)^{2}\right]\bar{c}\left(x_{1}\right)\gamma^{\mu}c\left(x_{2}\right)}
\]

\subsection{Four-quark vertex operators\label{sub:Four-quark-vertex-operators}}

The four-quark operators read as follows
\begin{eqnarray*}
\mathrm{Q_{1}=\left(\bar{c}_{a_{1}}b_{a_{2}}\right)_{V-A}\left(\bar{s}_{a_{2}}c_{a_{1}}\right)_{V-A}} & \qquad & Q_{4}=\left(\bar{s}_{a_{1}}b_{a_{2}}\right)_{V-A}\left(\bar{c}_{a_{2}}c_{a_{1}}\right)_{V-A}\\
\mathrm{\mathrm{Q_{2}=\left(\bar{c}_{a_{1}}b_{a_{1}}\right)_{V-A}\left(\bar{s}_{a_{2}}c_{a_{2}}\right)_{V-A}}} & \mathrm{\qquad} & \mathrm{Q_{5}=\left(\bar{s}_{a_{1}}b_{a_{1}}\right)_{V-A}\left(\bar{c}_{a_{2}}c_{a_{2}}\right)_{V+A}}\\
\mathrm{Q_{3}=\left(\bar{s}_{a_{1}}b_{a_{1}}\right)_{V-A}\left(\bar{c}_{a_{2}}c_{a_{2}}\right)_{V-A}} & \qquad & Q_{6}=\left(\bar{s}_{a_{1}}b_{a_{2}}\right)_{V-A}\left(\bar{c}_{a_{2}}c_{a_{1}}\right)_{V+A}
\end{eqnarray*}
with
\[
\mathrm{\left(\bar{\psi}\psi\right)_{V-A}=\bar{\psi}O^{\mu}\psi,\; O^{\mu}=\gamma^{\mu}\left(1-\gamma^{5}\right)\qquad\left(\bar{\psi}\psi\right)_{V+A}=\bar{\psi}O_{+}^{\mu}\psi,\; O_{+}^{\mu}=\gamma^{\mu}\left(1+\gamma^{5}\right).}
\]

\subsection{Some elements on factorization\label{sub:Some-elements-of}}

For the $\eta$ particle and the operator $\mathrm{Q_{1}}$, the time-product
matrix element is written as
\[
\mathrm{\left\langle T\left\{ \left[\bar{b}\left(x_{1}^{B}\right)i\gamma^{5}s\left(x_{2}^{B}\right)\right]C_{1}Q_{1}\left(x^{w}\right)\left[\bar{s}\left(x_{1}^{\eta}\right)i\gamma^{5}s\left(x_{2}^{\eta}\right)\right]\left[\bar{c}\left(x_{1}^{\psi}\right)\hat{\epsilon}_{\psi}c\left(x_{2}^{\psi}\right)\right]\right\} \right\rangle _{0}.}
\]
This, after evaluating the contractions and color indices, leads to
\begin{eqnarray*}
\mathrm{\left\langle T\left\{ \ldots Q_{1}\ldots\right\} \right\rangle _{0}} & \mathrm{=} & \mathrm{-C_{1}N_{c}^{2}\epsilon_{\psi,\nu}\times Tr\left[S_{b}\left(x^{w}-x_{1}^{B}\right)\gamma^{5}S_{s}\left(x_{2}^{B}-x_{1}^{\eta}\right)\gamma^{5}S_{s}\left(x_{2}^{\eta}-x^{w}\right)O^{\mu}\right]}\\
 &  & \mathrm{\times Tr\left[S_{c}\left(x^{w}-x_{1}^{\psi}\right)\gamma^{\nu}S_{c}\left(x_{2}^{\psi}-x^{w}\right)O^{\mu}\right],}
\end{eqnarray*}
where $\mathrm{S}$ is a propagator, index $\mathrm{w}$ refers to
the position of the four-quark interaction and $\mathrm{\epsilon_{\psi,\nu}}$
is the $\mathrm{J/\psi}$ polarization vector. The structure of the
expression makes visible the factorization into a ``$\mathrm{B_{s}-\eta}$''
and ``$\mathrm{J/\psi}$'' part. When actually comparing the two
parts to the expressions from \cite{Ivanov:2011aa}, one recognizes
the form factor and the decay constant. Situation is analogical for
$\eta'$ and other operators.

Further, very simple relations exist between the matrix elements for
different operators. One has
\[
\left\langle T\left\{ \ldots Q_{2,4,6}\ldots\right\} \right\rangle _{0}=\frac{1}{N_{c}}\left\langle T\left\{ \ldots Q_{1,3,5}\ldots\right\} \right\rangle _{0}.
\]

\bibliographystyle{apsrev}
\bibliography{BsDecay_1col}

\end{document}